\documentclass[10pt]{article}
\usepackage{citesort,epsfig}
\begin{document}

\title{Phase behaviour of a model of
colloidal particles with a fluctuating internal state}

\author{{\bf Richard P. Sear}\\
~\\
Department of Physics, University of Surrey\\
Guildford, Surrey GU2 5XH, United Kingdom\\
email: r.sear@surrey.ac.uk}
\date{}

\maketitle

\begin{abstract}
Colloidal particles are not simple rigid particles, in general
an isolated particle is a system with many degrees of freedom
in its own right, e.g., the counterions around a charged colloidal
particle.
The behaviour of model colloidal particles, with a simple
phenomenological model to account for these degrees of freedom, is
studied. It is found that the interaction between the particles
is not pairwise additive. It is even possible that the interaction
between a triplet
of particles is attractive while the pair interaction is
repulsive. When this is so the liquid phase is either stable
only in a small region of the phase diagram or absent altogether.
\end{abstract}

%\newpage
\section{Introduction}

Theories for the behaviour of systems of particles
usually apply to a model for the interaction between particles in
which the energy of interaction is pairwise additive.
The Hamiltonian is of the form
\begin{equation}
H({\bf r}^N) = \sum_{i,j=1}^{N~~'}u(r_{ij}),
\label{simplepot}
\end{equation}
a sum of spherically-symmetric pairwise-additive
potentials $u(r)$. ${\bf r}^N$ denotes the coordinates of the $N$ particles
and $r_{ij}$ is the scalar separation of particles $i$ and $j$.
But experiments are almost never on simple rigid
particles. For example, the particles in a colloidal suspension are not
simple rigid particles. Generally they
are stabilised in the suspension either by being charged, in which case they
are surrounded by a cloud of counterions, or by having
short polymers grafted to their surface \cite{russel}.
In either case we have not a rigid particle but a system with many
degrees of freedom, e.g., the positions of the counterions,
which fluctuate. Each particle is a system, with a free energy,
susceptibilities etc., in its own right. Here we will treat
systems of particles, each
of which is a weakly fluctuating statistical mechanical system
in its own right. Each particle will have a Landau-like free energy
which couples to that of neighbouring particles. We will use
the simplest possible form of this free energy, and
show exactly and analytically that it leads to many-body
attractions. These cannot be expressed as a Hamiltonian with
the form of Eq. (\ref{simplepot}).

This work is partly inspired by recent experiments on highly charged
colloidal particles under conditions of minimal amounts of added salt.
The potential of mean force between an {\it isolated} pair of
colloidal particles with minimal salt concentrations
has been measured and is purely repulsive \cite{crocker96}.
Yet the particles form crystallites which appear
to be metastable at close to zero osmotic pressure \cite{ito94}; this
is very hard to explain unless there is some sort
of cohesive attraction between the particles in the crystallites.
It does not seem possible to explain these findings
with a Hamiltonian which depends only on the coordinates of the
colloidal particles and has the conventional form of
Eq. (\ref{simplepot}). Here we develop
a simple phenomenological theory for colloidal particles which treats each
particle as fluctuating statistical mechanical system. 
See Refs.
\cite{roij98,roij99,denton99,levin98,levin99,warren,schmitz99,tehver99}
for recent theoretical work on understanding this
behaviour starting from a Hamiltonian which explicitly
includes the counterions and the electrostatic interactions.
The theory is phenomenological as we simply {\it assume} that
the state of a particle can be described by a single coarse-grained
scalar variable, we do not derive this.
This scalar
variable fluctuates and these fluctuations are perturbed
by the presence of another particle nearby.
The perturbations due to the particles which surround any
given particle add up, meaning that the state of a particle
changes with the number of its neighbours.
Essentially, the more particles that surround any given particle the
more the particle's fluctuations are biased towards
values that minimise the interaction free energy and so the more attractive
is the interaction between the given particle and all the surrounding
particles.
It is possible for two of our model particles to repel each other but
for particles in clusters of more than two particles to attract each other.
This can lead to coexistence
between a dilute fluid and a dense crystal phase even when a pair
of particles repel each other.
We will show that generally if the fluctuations can be described
by a scalar variable then the interaction between a pair
of particles within a cluster of several particles is more attractive than
between an isolated pair.

In the next section we define our model. In section 3 we show
that it exactly corresponds to a potential between structureless
particle which contains both pair and triplet terms. Then in section 4
we apply a perturbation theory to obtain an approximation
to the free energy of both the fluid and crystalline phases,
which is accurate for long-ranged interactions between the
fluctuating internal states. We show results for the phase
behaviour and for the zero wavevector structure factor in section 5.
Section 6 is a conclusion.

\section{Model}

Our model particles have an internal state specified
by a single scalar variable. The value of the variable for a single
isolated particle fluctuates weakly.
Essentially, we view it as
a coarse-grained variable \cite{chaikin}
obtained by averaging over some large number
of degrees of freedom associated with each particle.
An outline of how our mesoscopic Hamiltonian may be derived
from a microscopic one is given in the Appendix.
The interaction between
a pair of the particles depends on the value of this variable.
Thus, when two particles interact the mean values of the variables
on the two particles will change; the particles  `polarise'
each other, the interaction between them biases both internal variables
towards values for which their interaction free energy is low.
We included polarise in quotes because our model is phenomenological
not electrostatic; cf Ref. \cite{levin99} which considers
the polarisation of one a charged colloidal particle by another.

The effective Hamiltonian $H$ of the system of particles has two parts:
one part independent of the $s$ variables, $U$, and
the other 
is the free energy of the particles as a function
of the $N$ internal variables,
$F_N$,
\begin{equation}
H({\bf r}^N, s^N) = U({\bf r}^N) + F_N({\bf r}^N, s^N),
\label{hamil}
\end{equation}
where ${\bf r}^N$ and $s^N$ symbolise the centre-of-mass and
internal-variable coordinates, respectively, of all $N$
particles.

We need not specify $U$ beyond saying that it should be such
that the system has a well-defined thermodynamic limit \cite{ruelle64}.
Later on we will set $U$ to be the sum of
hard-sphere repulsions between pairs of the particles,
\begin{equation}
U({\bf r}^N) =\frac{1}{2} \sum_{i,j=1}^{N~~'}u_{hs}(r_{ij}),
\label{hs}
\end{equation}
where $u_{hs}$ is the hard sphere potential
\begin{equation}
u_{hs}(r)=\left\{
\begin{array}{ll}
0 & r\ge \sigma\\
\infty & r<\sigma
\end{array}\right. .
\end{equation}
$r$ is the separation of the centres of the interacting particles.
The dash over the sum in Eq. (\ref{hs}) means that the sum
is only over those terms for which $j\ne i$. Note that not all
the double and triple sums below will have a dash, in the undashed
ones terms with equal subscripts are summed over.

Each particle has a dimensionless internal variable $s$.
Now, we assume that the coupling between these variables on
different particles is pairwise additive and spherically symmetric.
Then the free energy $F_N$ is,
\begin{equation}
F_N({\bf r}^N,s^N)=\sum_{i=1}^Nf^{(1)}(s_i)
+\frac{1}{2}\sum_{i,j=1}^{N~~'}
f^{(2)}(s_i,s_j,r_{ij}).
\label{hstart}
\end{equation}
$f^{(1)}$ is the free energy of a single isolated particle
and is a function only of its $s$ variable.
$f^{(2)}$ is the
difference in free energy between an isolated pair of particles
and two isolated particles,
and is a function only of the two $s$ variables and
the magnitude of the separation of the two particles.
An isolated particle is a particle far from any other particle,
an isolated pair of particles is a pair of particles far from
any other particle.
For weak fluctuations $s$ is so we
Taylor expand $f^{(1)}$ and $f^{(2)}$
\begin{equation}
f^{(1)}(s)=\alpha s^2+\alpha_3s^3+\alpha_4s^4+\cdots,
\label{f1}
\end{equation}
where the linear term is missing as the variable $s$ is
defined so that if a particle is isolated its mean value is zero,
and
\begin{equation}
f^{(2)}(s,s',r)=
\phi_0(r)+\phi_1(r)(s_i+s')+
\phi_2(r)(s^2+s'^2)+\phi_2^x(r)ss'+\cdots.
\label{f2}
\end{equation}
The coefficients $\alpha$, $\alpha_3$ etc. are
derivatives of $f^{(1)}$
\begin{equation}
\alpha=\frac{1}{2}\frac{{\rm d}^2 f^{(1)}(s)}{{\rm d} s^2},
\end{equation}
and $\alpha_3$ is $1/6$ times the third derivative etc..
Similarly, the coefficients $\phi_0(r)$, $\phi_1(r)$ etc. are
derivatives of $f^{(2)}$ at a fixed separation of the particles.
$\phi_0$ is the zeroth derivative, $\phi_1$ is the first derivative
with respect to either of the two $s$ variables,
\begin{equation}
\phi_1(r)=\left(\frac{\partial f^{(2)}(s,s',r)}
{\partial s}\right)_{s',r},
\end{equation}
and similarly for the higher derivatives $\phi_2$ etc..

Our Taylor expansions are quite general for particles which have a state
which can be described by a single weakly fluctuating
scalar variable and in which
the interaction between two particles can be expressed in terms 
of this variable. If the variable is a vector or a
tensor not a scalar then clearly there will be expressions analogous to
that of Eq. (\ref{hstart})
but vectorial or tensorial variables
will in general lead to results very different from those
found here.

We truncate the Taylor expansions, Eqs. (\ref{f1}) and (\ref{f2}), after
their lowest nontrivial terms and substitute
the resulting expansions into Eq. (\ref{hstart}) to obtain
\begin{equation}
F_N({\bf r}^N,s^N) = \sum_{i=1}^N \alpha s_i^2+
\frac{1}{2}\sum_{i,j=1}^{N~~'}\left[
\phi_0(r_{ij})+\phi_1(r_{ij}) \left(s_i+s_j\right)\right].
\label{hf}
\end{equation}
The inverse susceptibility, $\alpha$, and
the zeroth, $\phi_0$, and first, $\phi_1$,
order coefficients all have the dimensions of energy.
So, for our Hamiltonian, Eqs. (\ref{hamil}) and (\ref{hf}),
to describe a system of particles
accurately we require:
(1) that the fluctuations in $s$ of an isolated
particle be sufficiently small that the
higher order terms in $f^{(1)}$ may be neglected,
(2) that the interactions between the
particles can be described as a sum over interactions of pairs,
(3) that this interaction is spherically symmetric,
(4) that this interaction can be accurately described by
function of a single scalar, coarse-grained, variable,
and
(5) that the interactions
between fluctuations are sufficiently weak that
the higher terms in $f^{(1)}$ and $f^{(2)}$ are small.

For noninteracting particles the second sum of Eq. (\ref{hf}) may be
neglected and $F_N$ is just a sum of independent quadratic
terms. The distribution of the $s$'s is then Gaussian,
which is correct for small fluctuations of a coarse-grained
variable \cite{landau,chaikin}.
$\alpha$ is an inverse susceptibility of an isolated particle,
the smaller it is the larger are the
fluctuations.
The interaction between a pair of particles is expressed as
a Taylor expansion in the two $s$'s truncated after the
linear term.
One particle feels an interaction due to another nearby particle,
which couples to its order parameter $s$ with
a strength $\phi_1(r)$.
The $s$-dependence of the free
energy of Eq. (\ref{hf}) is what we would expect if the
particles were weakly interacting, weakly fluctuating thermodynamic systems.

\section{Exact theory}

We start from the configurational integral $Z_N$ for $N$
particles in a volume $V$ and at a temperature $T$
\begin{equation}
Z_N=\int {\rm d}{\bf r}^N{\rm d}s^N\exp(-H({\bf r}^N,s^N)/kT),
\label{zn}
\end{equation}
with the Hamiltonian given by Eqs. (\ref{hamil}) and
(\ref{hf}). The Helmholtz free energy $A$ is then
\begin{equation}
A=-kT\ln\left(Z_N\Lambda^N/N!\right),
\end{equation}
where $\Lambda$ derives from the integration over the momenta.
Using Eqs. (\ref{hamil}) and (\ref{hf}) in Eq. (\ref{zn}),
\begin{eqnarray}
Z_N&=&\int {\rm d}{\bf r}^N\exp(-U({\bf r}^N)/kT)
\int {\rm d}s^N
\exp\left(- \sum_{i=1}^N (\alpha/kT) s_i^2-
\frac{1}{2}\sum_{i,j=1}^{N~~'}\left[
\phi_0(r_{ij})/kT+\phi_1(r_{ij})\left(s_i+s_j\right)/kT\right]\right)
\nonumber\\
&=&\int {\rm d}{\bf r}^N
\exp\left(-\frac{U({\bf r}^N)}{kT}-
\frac{1}{2}\sum_{i,j=1}^{N~~'}\frac{\phi_0(r_{ij})}{kT}\right)
\prod_{i=1}^N\left\{\int_{-\infty}^{\infty}{\rm d}s_i
\exp\left(-\frac{\alpha}{kT} s_i^2-
\sum_{j=1~j\ne i}^N
\frac{\phi_1(r_{ij})}{kT}s_i\right)\right\},
\end{eqnarray}
where to obtain the second line we
expressed the integral over the $s$ variables
as a product of integrations over each one and then grouped all the terms
which depend on each $s_i$ together.
Each integration over an $s$ variable is independent
and can be done easily
\begin{eqnarray}
Z_N&=&\int {\rm d}{\bf r}^N
\exp\left(-\frac{U({\bf r}^N)}{kT}-
\frac{1}{2}\sum_{i,j=1}^{N~~'}\frac{\phi_0(r_{ij})}{kT}\right)
\prod_{i=1}^N\left\{\left(\frac{\pi kT}{\alpha}\right)^{1/2}
\exp\left(\frac{1}{4\alpha kT}\left[\sum_{j=1~j\ne i}^N
\phi_1(r_{ij})\right]^2\right)\right\}
\nonumber\\
&=& \left(\frac{\pi kT}{\alpha}\right)^{N/2}
\int{\rm d}{\bf r}^N
\exp\left(-\frac{U({\bf r}^N)}{kT}-
\frac{1}{2}\sum_{i,j=1}^{N~~'}\frac{\phi_0(r_{ij})}{kT}\right)
\prod_{i=1}^N\left\{
\exp\left(\frac{1}{4\alpha kT}\sum_{j=1~j\ne i}^N\sum_{k=1~k\ne i}^N
\phi_1(r_{ij})\phi_1(r_{ik})\right)\right\}
\nonumber\\
&=&
\left(\frac{\pi kT}{\alpha}\right)^{N/2}
\int {\rm d}{\bf r}^N
\exp\left(-\frac{U({\bf r}^N)}{kT}-
\frac{1}{2}\sum_{i,j=1}^{N~~'}\frac{\phi_0(r_{ij})}{kT}
+\frac{1}{4\alpha kT}
\sum_{i,j,k=1~j\ne i~k\ne i}^N
\phi_1(r_{ij})\phi_1(r_{ik})\right),
\label{pfr1}
\end{eqnarray}
where to obtain the second from the first
line we expressed the square of the sum as a double sum. To
obtain the third line we converted the product of exponentials
to the exponential of a sum. The factor in front of the
integration is of course very familiar: it is just the partition
function of $N$ independent simple harmonic oscillators. It does
not depend on density and so has no effect on the phase behaviour.
Note that in the triple sum although neither $j$ nor $k$ can be equal
to $i$, $j$ can be equal to $k$. The restrictions on $j$ and $k$
derive from the fact that a particle cannot interact with itself,
which would correspond to $j,k=i$ but as both
$j$ and $k$ in the triple sum come from the square of a single
sum they are in effect from the same interaction and therefore
are allowed to be equal. We can rewrite Eq. (\ref{pfr1}) by
extracting the $j=k$ terms from the triple sum,
\begin{equation}
Z_N=\left(\frac{\pi kT}{\alpha}\right)^{N/2}
\int {\rm d}{\bf r}^N
\exp\left(-\frac{U({\bf r}^N)}{kT}-
\frac{1}{2}\sum_{i,j=1}^{N~~'}\left[
\frac{\phi_0(r_{ij})}{kT}
-\frac{\phi_1(r_{ij})^2}{2\alpha kT}\right]
+\frac{1}{4\alpha kT}
\sum_{i,j,k=1}^{N~~'}
\phi_1(r_{ij})\phi_1(r_{ik})\right),
\nonumber\\
\label{pfr}
\end{equation}
where the triple sum has a dash to indicate that terms in which
any of the three subscripts are the same are excluded.
The configurational integral,
Eq. (\ref{pfr}), is an integral only over the positions of the $N$ particles.
Neglecting the irrelevant (for the phase behaviour) prefactor
it is nothing but the configurational integral
of the Hamiltonian, $H_{eff}$,
\begin{equation}
H_{eff}({\bf r}^N)=U({\bf r}^N)+
\frac{1}{2}\sum_{i,j=1}^{N~~'}\left[
\phi_0(r_{ij})
- \frac{\phi_1(r_{ij})^2}{2\alpha}\right]
-\frac{1}{6}
\sum_{i,j,k=1}^{N~~'}\frac{\left[
\phi_1(r_{ij})\phi_1(r_{ik})+
\phi_1(r_{ij})\phi_1(r_{jk})+
\phi_1(r_{ik})\phi_1(r_{jk})\right]}{2\alpha},
\nonumber\\
\label{heff}
\end{equation}
where we have rewritten the summand of the triple sum to make
it symmetric with respect to the three indices.
The phase behaviour of our model particles will be identical
to that of structureless, spherically symmetric particles
with interactions described by the Hamiltonian $H_{eff}$ of Eq. (\ref{heff}).
The Hamiltonian is that of a triplet or 3-body potential; the first two sums
are conventional sums over pair potentials but the last sum is
over a 3-body potential. We started with a pair potential which
depended on the $s$ variables as well as the positions,
integration
over the $s$ variables resulted in a potential which
depends only on the positions but is no longer a simple pair potential.

The effective Hamiltonian, $H_{eff}$, can lead
to behaviour qualitatively different from that of a Hamiltonian
which is a simple sum over a pair potential. To see this we
will compare the interaction between a pair of particles
to that between a triplet of particles. For a
pair of particles, 1 and 2, a distance $r$ apart,
the interaction is from Eq. (\ref{heff}),
\begin{equation}
H_{eff}({\bf r}^2)=U({\bf r}^2)+
\phi_0(r)-\frac{1}{2}\frac{\phi_1(r)^2}{\alpha}.
\label{avef2}
\end{equation}
Note that the contribution due to the fluctuations, the part inversely
proportional to $\alpha$ is always
negative regardless of the sign of $\phi_1$,
fluctuations always produce an effective attraction.
Now, consider 3 particles, at the
corners of an equilateral triangle of side $r$ for simplicity.
We denote the set of 3 coordinates marking the corners
of an equilateral triangle by ${\bf r}_e^3$.
The interaction is, from Eq. (\ref{heff}),
\begin{equation}
H_{eff}({\bf r}^3_e)=U({\bf r}^3_e)
+3\phi_0(r)
-3\frac{\phi_1(r)^2}{\alpha}.
\label{aveu3}
\end{equation}
If we compare this
with the interaction of a pair, Eq. (\ref{avef2}), we see that the ratio of the
$\phi_1^2/\alpha$ to $\phi_0$ term has doubled.
For a pairwise additive potential, the interaction
free energy would be simply three times Eq. (\ref{avef2}):
$3\phi_0-(3/2)\phi_1^2/\alpha$.
The relative contribution from
the fluctuations in the $s$ variables has doubled.
This contribution is always attractive so the free energy
of attraction is always more negative than for a pairwise additive
potential.
This is a general result for a weakly fluctuating scalar
variable.
Our result for the interaction between three particles
may be compared with the more microscopic work on the interaction
between three particles of L\"{o}wen and coworkers. They considered
triplet interactions between charged colloidal particles
\cite{lowen98} and between star polymers \cite{ferber00}.

When $\phi_0(r)>0$ and
$\phi_0(r)<\phi_1(r)^2/\alpha<2\phi_0(r)$, the interaction minus
the part from $U$ is repulsive, i.e., greater then zero, for
a pair $r$ apart but attractive for a triplet at the corners of an
equilateral triangle of side $r$. If $U$ is some short-range repulsion,
e.g., a hard sphere repulsion, which is zero at some sufficiently
large value of $r$, then
it is possible for a pair of our model colloidal particles
to repel each other, but for a triplet of them to
attract.
For other arrangements of 3 particles the sign of the
interaction free energy will depend on the values
of the 3 separations of the centres of the particles and on the
distance dependence of $\phi_0$ and $\phi_1$.
However, this should not obscure the basic fact that
the interaction free energy of 3 particles can be negative when the
interaction of 2 particles is positive.
For 4 particles at the corners of a tetrahedron the interaction
minus the part from $U$ is
$6\phi_0-9\phi_1^2/\alpha$
Again, it is possible to
obtain negative interaction free energies even when the interaction
free energy of a pair is always positive. Indeed it is possible to obtain
negative free energies even when they are positive for three particles
at the corners of an equilateral triangle.

%For a configuration of the particles, ${\bf r}^N$, we can determine
%the mean value of the $s$ variable of the $i$th particle,
%${\overline s_i}({\bf r}^N)$, by averaging over $s_i$ with
%a weight containing the $s_i$ dependent parts of our Hamiltonian,
%Eq. (\ref{hf}),
%\begin{eqnarray}
%<s_i({\bf r}^N)>&=&\frac{
%\int_{-\infty}^{\infty}{\rm d}s_i s_i
%\exp\left(-\frac{\alpha}{kT}s_i^2-\sum_{j=1~j\ne i}^N
%\frac{\phi_1(r_{ij})}{kT}s_i\right)}
%{\int_{-\infty}^{\infty}{\rm d}s_i
%\exp\left(-\frac{\alpha}{kT}s_i^2-\sum_{j=1~j\ne i}^N
%\frac{\phi_1(r_{ij})}{kT}s_i\right)}\\
%&=&-\frac{1}{2\alpha}\sum_{j=1~j\ne i}^N\phi_1(r_{ij}).
%\label{aves}
%\end{eqnarray}
%The mean value of an $s$ variable is a linear function, the contribution
%of each interaction with one of the surrounding particles is independent.

\section{Mean-field free energy}

In order to able to keep the theory simple
we now specialise to an interaction between the internal variables
which is both weak and long-ranged. We
also set $U$ to be the hard-sphere interaction.
Weak in the sense that
$|\phi_0(r)|,|\phi_1(r)\phi_1(r')/\alpha|\ll kT$
for all values of $r$, $r'$, and long-ranged as
both $\phi_0(r)$ and $\phi_1(r)$ decay to zero over a
characteristic length scale which is much longer than the hard-sphere
diameter $\sigma$. So the interaction Hamiltonian, Eq. (\ref{heff}),
which is a function of the positions only,
consists of a hard-sphere interaction, $U$, and a weak
long-ranged interaction which has two parts: a pair potential
and a triplet potential. Weak but long-range many-body interactions have been
considered by the author in Ref. \cite{sear00}.

We view the long-range part of the interactions as a perturbation
\cite{hansen86}.
The free energy is expressed as that of hard spheres, $A_{hs}$,
plus the difference in free energy between a system of our particles
and that of hard spheres, $\Delta A$,
\begin{equation}
A = A_{hs} + \Delta A.
\label{free}
\end{equation}
For the hard sphere free energy we use the
expression of Carnahan and Starling \cite{carnahan69} for
the fluid phase and the fit to simulation data of Hall \cite{hall72}
for the solid phase.
We approximate $\Delta A$ by an average of the perturbing part
of the Hamiltonian, $H_{eff}-U$,
\begin{equation}
\Delta A=\frac{\int {\rm d}{\bf r}^N
\left(H_{eff}({\bf r}^N)-U({\bf r}^N)\right) \exp\left(-U({\bf r}^N)/kT\right)}
{Z_{hs}},
\label{da}
\end{equation}
where $Z_{hs}$ is the configurational integral for $N$ hard spheres.

Substituting Eq. (\ref{heff}) into Eq. (\ref{da}),
\begin{equation}
\Delta A=Z_{hs}^{-1}\int {\rm d}{\bf r}^N
\exp\left(-U({\bf r}^N)/kT\right)
\left\{
\frac{1}{2}N(N-1)
\left(\phi_0(r_{12})-\frac{\phi_1(r_{12})^2}{2\alpha}\right)-
\frac{1}{6}(N-1)(N-2)
\frac{3\phi_1(r_{12})\phi_1(r_{13})}{2\alpha}\right\}.
\nonumber\\
\label{udef2}
\end{equation}
The $n$-particle density of hard spheres, $\rho^{(n)}_{hs}$,
and their $n$-particle distribution function, $g^{(n)}_{hs}$,
are defined as \cite{hansen86}
\begin{eqnarray}
\rho^{(n)}_{hs}({\bf r}^n)&=&
\left(\prod_{i=1}^n\rho^{(1)}_{hs}
({\bf r}_i)\right) g^{(n)}_{hs}({\bf r}^n)\nonumber\\
&=&\frac{N!}{(N-n)!}\frac{\int
\exp\left[-\beta U({\bf r}^N)\right]
{\rm d}{\bf r}^{N-n}}{Z_{hs}}.
\label{gdef}
\end{eqnarray}
The 1-particle density, $\rho^{(1)}_{hs}({\bf r})$,
is not assumed to be uniform
so that the theory applies to crystalline as well as fluid phases.
Using, Eq. (\ref{gdef}) in Eq. (\ref{udef2}), we obtain
\begin{equation}
\Delta A=
\frac{1}{2}\int{\rm d}{\bf r}^2
\rho^{(2)}_{hs}({\bf r}_1,{\bf r}_2)
\left(\phi_0(r_{12})-\frac{\phi_1(r_{12})^2}{2\alpha}\right)
-\frac{1}{4\alpha}\int{\rm d}{\bf r}^3
\rho^{(3)}_{hs}({\bf r}_1,{\bf r}_2,{\bf r}_3)
\phi_1(r_{12}) \phi_1(r_{13}).
\nonumber\\
\label{genda}
\end{equation}

As $\phi_0,\phi_1$ decay to zero only
when the separations of the particles are much larger than $\sigma$,
the integrals in Eq. (\ref{genda}) are
dominated by configurations when the spheres are far apart, i.e.,
where the pair separations are much larger than $\sigma$.
In a fluid the one particle density is a constant, $\rho^{(1)}=\rho$,
and at separations large with respect to $\sigma$ the distribution
function is close to one, $g^{(n)}_{hs}\simeq 1$. Thus in the fluid phase
the integrands of Eq. (\ref{genda}) can be simply approximated by
$\rho^2(\phi_0(r_{12})-\phi_1(r_{12})^2/2\alpha)$
and $\rho^3\phi_1(r_{12})\phi_1(r_{13})$.
In a crystalline phase, although there are long-range
correlations in $\rho^{(n)}_{hs}$ the one particle
density, $\rho^{(1)}_{hs}$, averaged over a unit cell is
just $\rho$.
The attractive interaction between particles has a range much larger than
the lattice constant of the lattice (a little larger than $\sigma$) and so
as $\phi_0$ and $\phi_1$ vary
little across a unit cell we can regard $\rho^{(1)}_{hs}({\bf r})$
as approximately constant at its average value, $\rho$.
Similarly, as we change any one of the $n$ position vectors
upon which the $n$-body distribution function, $\rho^{(n)}_{hs}$, depends
the density oscillates rapidly over each unit cell but averages to
$\rho$. So, in the crystalline phase
as well as in the fluid phase we approximate $\rho^{(n)}_{hs}$ by $\rho^n$.
Then the integrands are the same as in a fluid phase. We have
\begin{equation}
\Delta A=
\frac{1}{2}V\rho^2\int_{\sigma}^{\infty}{\rm d}{\bf r}
\left(\phi_0(r)-\frac{\phi_1(r)^2}{2\alpha}\right)
-\frac{1}{4\alpha}V\rho^3\left(\int_{\sigma}^{\infty} {\rm d}{\bf r}
\phi_1(r)\right)^2.
\label{da2}
\end{equation}
Now the term coming from the square of $\phi_1$ is negligible if
$\phi_1$ is long-ranged. To see this consider an explicit
functional form for $\phi_1$. We choose a Kac potential,
\begin{equation}
\phi_1(r)=\epsilon\gamma^3\exp(-\gamma r/\sigma).
\label{kac}
\end{equation}
The range of this function is $\sigma/\gamma$, i.e., $\gamma^{-1}$
is a range in units of $\sigma$. Now the integral of Eq. (\ref{kac})
over 3-dimensional space is ${\cal O}(\epsilon\sigma^3)$;
essentially the integrand is of order $\epsilon\gamma^3$
within a volume of order $(\sigma/\gamma)^3$, and
negligible outside of this volume.
However, the integral
of the square of Eq. (\ref{kac}) over 3-dimensional space is
${\cal O}(\epsilon\sigma^3\gamma^3)$. For a long-range $\phi_1$,
the inverse range $\gamma\ll1$ and so the integral of the
square is negligible. We chose a specific functional form for
$\phi_1$ simply for clarity, our conclusion that the
integral of the square is negligible holds for any
long-range slowly decaying function.

So, in Eq. (\ref{da2}), we neglect the $\phi_1^2$ term,
integrate over
the remaining terms in Eq. (\ref{da2}) and obtain
\begin{equation}
\frac{\Delta A}{N}=\frac{1}{2}\rho\nu_0
-\frac{1}{4}\frac{(\rho\nu_1)^2}{\alpha},
\label{dares}
\end{equation}
where $\nu_0$ and $\nu_1$ are the integrals
\begin{equation}
\int_\sigma^{\infty}{\rm d}{\bf r}\phi_i(r)=\nu_i, ~~~~i=0,1.
\end{equation}
The pressure $p$ can be easily derived from
the free energy, Eq. (\ref{dares}),
\begin{equation}
p=p_{hs}+\frac{1}{2}\rho^2\nu_0-\frac{1}{2}\frac{\rho^3\nu_1^2}{\alpha},
\label{press}
\end{equation}
where $p_{hs}$ is the pressure of hard spheres.
The contribution
from the fluctuations in the $s$ variables
to the pressure is cubic in the density as it must, it
is equivalent to a long-range 3-body attraction \cite{sear00}.
The chemical potential
$\mu$ is then simply given by $\mu=A/N+p/\rho$.
The pressure and chemical potential as functions of the temperature
and density allow us to calculate phase diagrams.

\section{Results}

The free energy, Eq. (\ref{dares}), depends on a single
parameter, the dimensionless ratio $R=\nu_0\alpha\sigma^3/\nu_1^2$.
It ranges from $-\infty$ to $+\infty$; the sign of $R$ is determined
by the sign of $\nu_0$ as $\alpha$ must be positive.
The larger the magnitude of $R$ the more dominant is the part of the
interaction potential which does not depend on $s$ and so the closer the
interaction is to a pairwise additive potential.
In the $R\rightarrow -\infty$ limit the interaction
is a simple pairwise additive attractive potential. Then
the free energy is just a modification
of the free energy of
van der Waals
\cite{hansen86,kampen64,lebowitz66,longuet64}.
The modification being
the replacement of his approximate free energy of hard spheres
by that of Carnahan and Starling \cite{carnahan69} in the fluid,
and that of Hall \cite{hall72} in the crystal.
In the $R\rightarrow +\infty$ limit the potential is again pairwise
additive but it is repulsive. A long-range repulsion tends
to cause not bulk vapour-liquid separation but microphase separation,
see Refs. \cite{lebowitz66,sear99}.
The $R\rightarrow\pm\infty$ limits result from
either the interaction between particles not depending on
the value of the $s$ variables, $\nu_1\rightarrow0$, or the
fluctuations of the $s$ variables tending to zero,
$\alpha\rightarrow\infty$.

When $R$ is finite then the interactions depend on the
fluctuations in $s$ and are no longer pairwise additive.
For $R<0$ the interactions are attractive even in the absence of 
fluctuations in $s$; the fluctuations in $s$ merely make the attractions
stronger and not pairwise additive. 
For $R>0$ the pairwise interactions are repulsive while the
3-body interactions from the fluctuations are, as always, attractive.
The part of the free energy due to fluctuations, the last
term in Eq. (\ref{dares}),
is one power higher in the density than
the part independent of the fluctuations, the second
in Eq. (\ref{dares}). Thus, if $\nu_0>0$ the interactions are
always repulsive at sufficiently low densities but become
attractive, i.e., contribute a negative amount to the free energy
of the system, at a density $2R/\sigma^3$.

\begin{figure}
\caption{
The phase diagram of hard spheres plus a long-range pairwise
additive attraction, a van-der-Waals fluid.
$\nu_1=0$, $\nu_0<0$ and the reduced temperature $T^*=kT\sigma^3/|\nu_0|$.
The thick solid curves separate the one and two-phase regions.
The letters V, L and C denote the regions of the phase space
occupied by the vapour, liquid and crystalline phases, respectively.
The horizontal thin lines are tie lines connecting coexisting densities,
and the dotted line connects the three coexisting densities at the
triple point.
}
\label{fig1}
\begin{center}
\epsfig{file=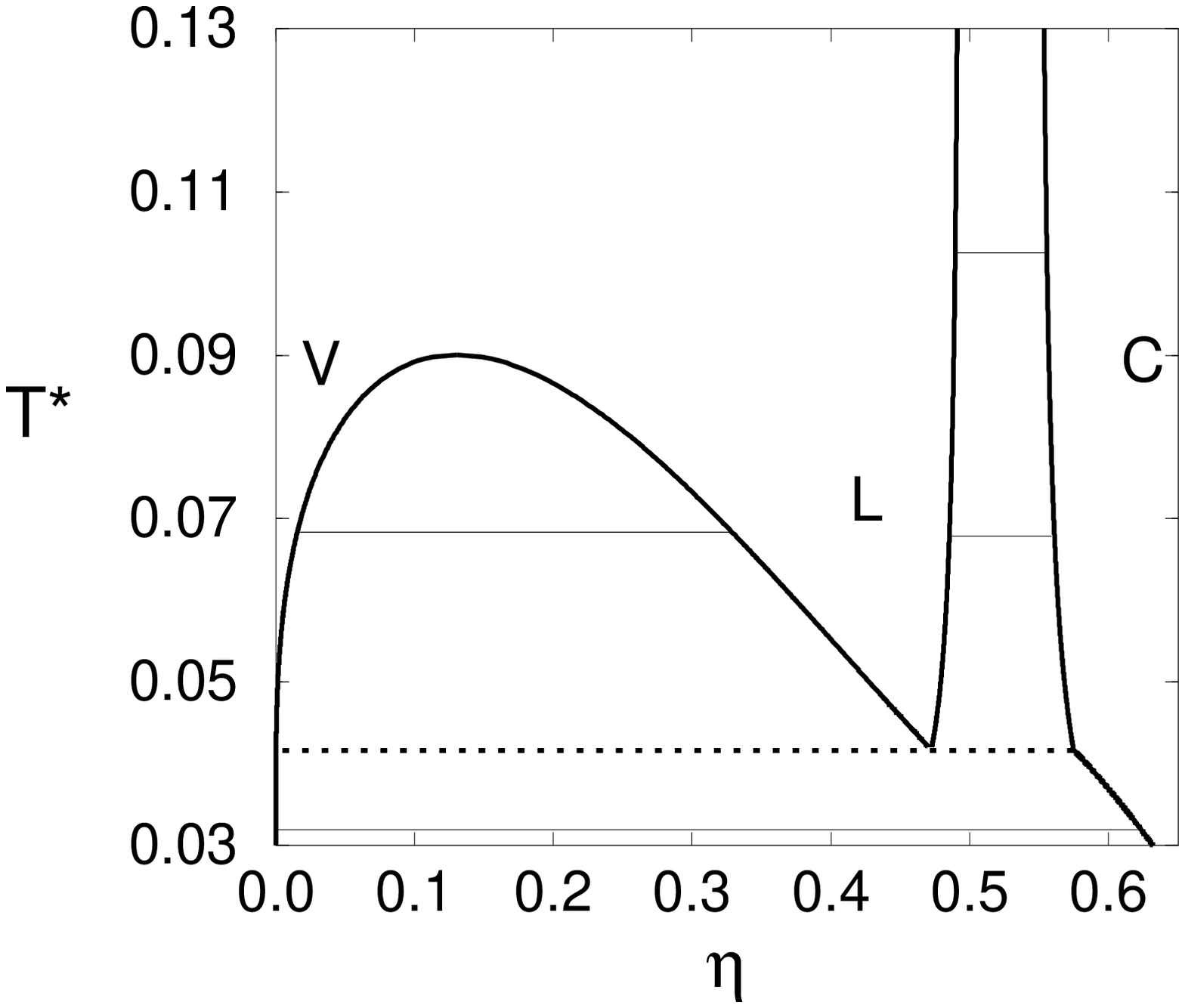,width=4.0in}
\end{center}
\end{figure}

We now discuss some example phase diagrams. The diagrams
are in the density-temperature plane. We use the reduced
density $\eta=(\pi/6)\rho\sigma^3$, which is equal to the
fraction of the volume occupied by the hard cores, and a reduced
temperature $T^*$ which is either $kT\sigma^3/|\nu_0|$
or $kT\alpha\sigma^6/\nu_1^2$.
For reference
we plot the familiar phase diagram of hard spheres with a long-range
attraction, often called the van-der-Waals
fluid \cite{kampen64,lebowitz66,longuet64}; the $R\rightarrow-\infty$ limit.
This has a free energy
given by Eqs. (\ref{free}) and (\ref{dares}) with $\nu_1=0$ and $\nu_0<0$.
The first condition means that the internal variables on different
particles do not couple and so the attraction is a simple
pair potential and the second condition makes the long-range interaction
attractive so there is a vapour-liquid transition, i.e., 
phase separation into two fluid phases of different densities.
The phase diagram is shown in Fig. \ref{fig1}; there is a large
temperature range over which there is stable vapour-liquid coexistence.
This is of course not new, in the $\nu_1=0$ limit our model
reduces to the most basic model of particles which form a liquid.

\begin{figure}
\caption{
The phase diagram of hard spheres plus a long-range interaction
proportional to the fluctuating internal variable $s$.
$\nu_0=0$, and the reduced temperature $T^*=kT\alpha\sigma^6/\nu_1^2$.
See caption to Fig. \ref{fig1} for the meaning of the letters, curves and
horizontal lines.
}
\label{fig3}
\begin{center}
\epsfig{file=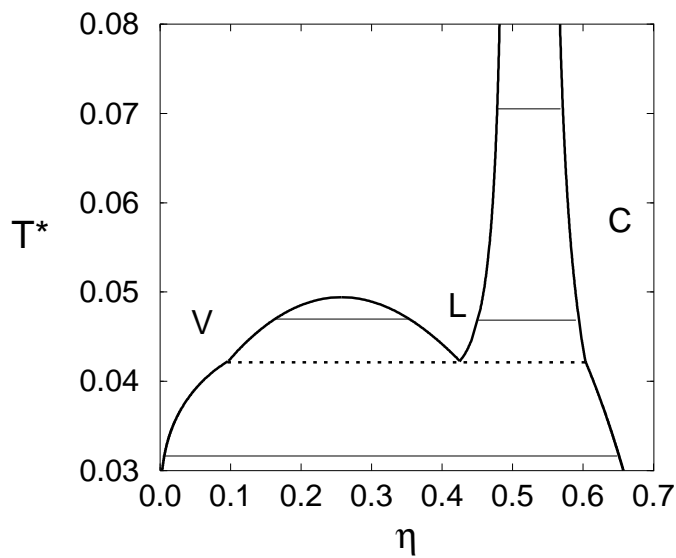,width=4.0in}
\end{center}
\end{figure}

In Fig. \ref{fig3}
we show the phase diagram of particles with $R=0$.
The long-range interaction
is purely proportional to the internal
variable $s$, i.e., $\nu_0=0$. There is vapour-liquid coexistence
over a range of temperatures but this range is much smaller than
for the simple pair potential of Fig. \ref{fig1}. Also, the density
at the critical point is higher; it is
almost double that in Fig. \ref{fig1}.
The interaction between $s$ variables is effectively
a long-range 3-body attraction, i.e., an attraction
which is van-der-Waals like except for the fact that it is
between triplets not pair of particles \cite{sear00}.
The phase diagram, Fig. \ref{fig3}, differs
from Fig. 2 of Ref. \cite{sear00} only in that
the temperature scale is different.

\begin{figure}
\caption{
The phase diagram of hard spheres plus a long-range
repulsion and a long-range attraction
proportional to the fluctuating internal variable $s$.
$R=-0.5$, and the reduced temperature
$T^*=kT\alpha\sigma^6/\nu_1^2$.
The letters F and C denote the regions of the phase space
occupied by the fluid and crystalline phases, respectively.
The dotted curve is the bulk free energies prediction of a
vapour-liquid--like transition within the fluid-crystal coexistence
region. 
See caption to Fig. \ref{fig1} for the meaning of the other curves and the
horizontal lines.
}
\label{fig4}
\begin{center}
\epsfig{file=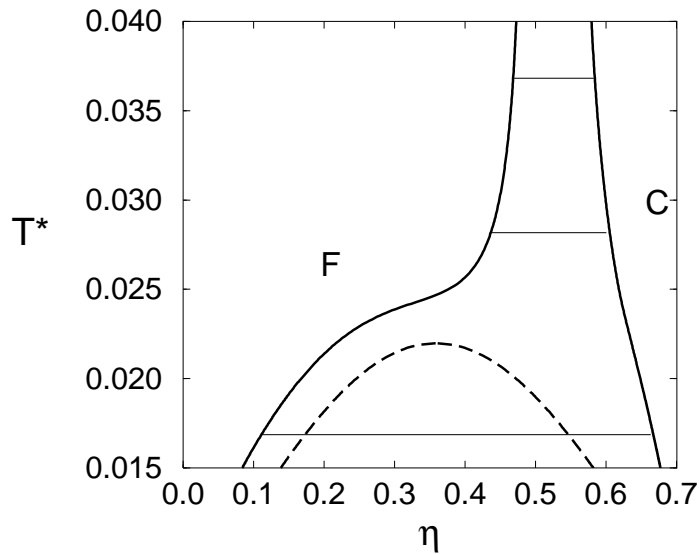,width=4.0in}
\end{center}
\end{figure}

In Fig. \ref{fig4} we show the phase diagram of particles with $R=2$.
There is no vapour-liquid coexistence at equilibrium, only
a fluid-crystal coexistence region which broadens dramatically
at low temperatures. The liquid phase has disappeared.
For other examples of liquid phases disappearing see Ref. \cite{faraday};
the most studied system in which the liquid disappears is that
of particles with a short-range, pairwise additive attraction
\cite{gast83,hagen94,daanoun94,dijkstra99}.
Within the fluid-crystal coexistence region the pressure and chemical
potential have van-der-Waals loops so our bulk free energy
predicts vapour-liquid coexistence within the fluid-crystal
coexistence region. Note that the density at the critical point
is even higher than in Fig. \ref{fig3}.

We have only considered fluid phases and crystalline phases.
However, sufficiently strong long-range
repulsions can transform a bulk phase separation into microphase
separation \cite{lebowitz66,sear99}. Essentially, microphase separation
occurs when the interaction between particles is repulsive
at the largest separations at which they interact separately
and the repulsion is sufficiently strong and long ranged.
Thus our present theory, which
neglects the possibility of microphase separation, should not be
used if the long-range interaction is predominantly
repulsive, i.e., if $\nu_0>0$, and either $\phi_0$ is longer
ranged than $\phi_1$, or $R\gg1$.

\subsection{The structure of the fluid phase}

We have shown that the phase behaviour of our particles can be
very different from that of a simple pair potential but what about the
structure? Of course the pair distribution function or
equivalently the structure factor \cite{hansen86} depends on the functional form
of the interactions, i.e., on the precise forms of the functions
$\phi_0(r)$ and $\phi_1(r)$. However, we have only considered
long-range interactions and there the bulk phase behaviour is insensitive
to the precise details of $\phi_0(r)$ and $\phi_1(r)$, it only
depends on their integrals $\nu_0$ and $\nu_1$. Also, long-range
interactions only affect the structure factor $S(q)$ at small wavevectors
$q$; small meaning of order the reciprocal of their range or less.
Thus, we will consider only the zero wavevector limit, $S(0)$, of the
structure factor. This is simply related to a thermodynamic
quantity, the isothermal compressibility $\chi_T$, by \cite {hansen86}
\begin{equation}
S(0)=\rho kT\chi_T,
\label{s0}
\end{equation}
where
\begin{equation}
\chi_T^{-1}=\rho\left(\frac{\partial p}{\partial \rho}\right)_{T}.
\label{chidef}
\end{equation}
Using Eq. (\ref{press}) for the pressure
\begin{equation}
\chi_T^{-1}=
\rho\left(\frac{\partial p_{hs}}{\partial \rho}\right)_{T}
+\rho^2\nu_0-\frac{3}{2}\frac{\rho^3\nu_1^2}{\alpha}.
\label{chi}
\end{equation}
Eqs. (\ref{s0}) and (\ref{chi}) give us the zero wavevector
structure factor of our particles. Below we will consider the density
dependence of $S(0)$. It will therefore be useful to obtain $S(0)$
as a density expansion. To do this we start by inserting the virial
expansion for the pressure into the definition of the
isothermal compressibility $\chi_T$, Eq. (\ref{chidef}).
Inserting this expansion into Eq. (\ref{s0}) for $S(0)$ yields
\begin{equation}
S(0)=1-2B_2\rho+{\cal O}({\rho^2}),
\label{virial}
\end{equation}
the initial slope of $S(0)$ is equal to minus twice the
second virial coefficient $B_2$. For our particles
\begin{equation}
B_2=\frac{2}{3}\pi\sigma^3-\frac{1}{2}\nu_0,
\label{b2}
\end{equation}
where the first term on the right hand side is the second virial
coefficient of hard spheres. Note that the fluctuations in the
$s$ variables do not contribute to the second virial
coefficient. This is only the case in the long-range interaction
limit. In section 3 and in Eq. (\ref{da2}) we found that in
general the fluctuations did contribute to the interaction between
pairs and hence to the second virial coefficient but that this
contribution was weaker than its contribution to the interaction
between larger numbers of particles and to the third virial coefficient.
The term in the free energy per unit volume
proportional to $\nu_1^2/\alpha$ which
comes from the fluctuations varies with density as $\rho^3$ and
so it contributes (only) to the third virial coefficient.

\begin{figure}
\caption{
The zero wavevector structure factor of our particles in the fluid
phase, $S(0)$, as a function of the volume fraction $\eta$.
The solid curve is for an attraction which does not depend on $s$;
the potential with the phase diagram Fig. \ref{fig1}.
$\nu_0/(\sigma^3kT)=-10$, which
is just above the critical temperature
The dashed curve is for an attraction which is
proportional to the fluctuating internal variable $s$;
the potential with the phase diagram Fig. \ref{fig3}.
$\nu_1^2/(\sigma^6kT\alpha)=19.4$, which
is again just above the critical temperature
}
\label{s0fig}
\begin{center}
\epsfig{file=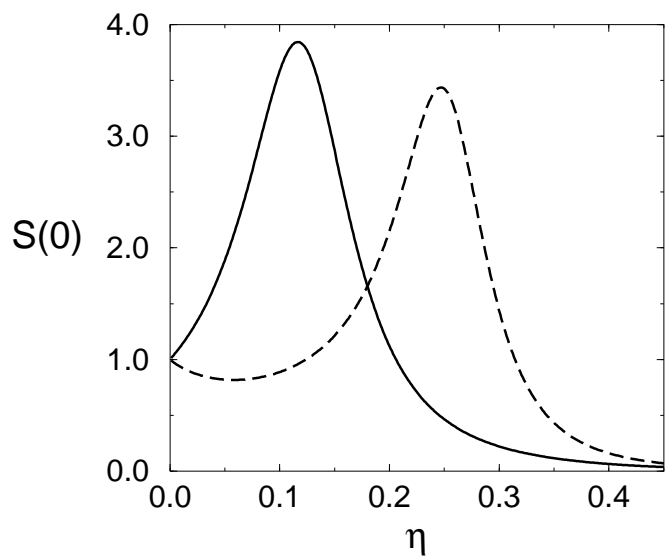,width=4.0in}
\end{center}
\end{figure}

In Fig. \ref{s0fig} we have plotted the zero wavevector structure factor
for a van-der-Waals fluid, hard spheres plus a long-range
pairwise additive attraction (the solid curve), and for an attraction which is
proportional to $s$ (the dashed curve). They are both at temperatures
just above the critical temperature. The peaks in $S(0)$ are due to
the nearby critical points, where $S(0)$ diverges of course.
The striking difference between the curves is their opposite slopes
at low densities. The limiting slope at vanishing densities is
equal to $-2B_2$, Eq. (\ref{virial}). For a van-der-Waals fluid
the second virial coefficient is negative at the critical temperature
so the slope of $S(0)$ is positive until above the critical density.
However, interactions between fluctuations do not contribute
to the second virial coefficient and so it is equal to that of
hard spheres which is positive. Thus the slope of $S(0)$ is negative
at low densities and $S(0)$ goes through a minimum below the critical density.

\section{Conclusion}

Particles which have a fluctuating internal state
interact in a way which is different from structureless particles.
The interaction is intrinsically many body and cannot be
described using a pair potential.
It is even possible for pairs of particles with a fluctuating
state to repel but larger clusters of particles to attract each other.
We have studied a rather generic model of a particle with
a fluctuating state.
Remarkably, because of the very simple form of our Hamiltonian's
dependence on the variables which describe the
states of the particles, see Eq. (\ref{hf}), we have been able
to integrate over these variables exactly and
analytically. The result is an interaction between
the particles which has both pair and triplet interactions,
Eq. (\ref{heff}).
Our model only included the leading order terms in Taylor
expansions for the free energy associated with the internal
states of the particles. It is possible to include higher
order terms although then the $s$ variables will have to
integrated over approximately or numerically,
alternatively computer simulation could be used. In any
case the effective potential will then contain not only
pair and triplet terms but terms of all orders, 4-body,
5-body, etc..

It is perhaps of interest to contrast the particles with
mesoscopic fluctuations studied here to the rods with
microscopic fluctuations studied by Ha and Liu \cite{ha},
see also Refs. \cite{kirkwood52,gronbech97,podgornik98}.
They studied the interactions between parallel rods and allowed
the charge density along each rod to fluctuate; if
a rod is along the $z$-axis then we can define a 1-dimensional charge
density $\rho(z)$, which will fluctuate. The Coulomb interaction
between charges will then couple the fluctuations in $\rho(z)$ in
nearby rods. Ha and Liu found that not only can this coupling
generate an attraction but that it is strongly nonpairwise
additive. In contrast with their model our is rather general
and simpler.
The results for our rather non-specific model
show clearly that these findings of Ha and Liu are rather generic.
The disadvantage of our model with
respect to that of Ha and Liu is that to relate it to
experiment requires determining or at least guessing
the appropriate values of the parameters of our
phenomenological interaction.

We hope that our model will be useful for determining generic
differences in behaviour between rigid particles interacting
via a pair potential and particles with a fluctuating state.
For example, although all our calculations have been for a
single component they are easily generalised to mixtures.
When this is done, it is straightforward to see that,
if the fluctuations between the internal states
of particles of different components are only weakly coupled then
this will tend to drive phase separation of these components.
However, differences between the inverse susceptibility, $\alpha$,
cannot drive phase separation. At least within our theory
which includes only leading order terms.

It is a pleasure to acknowledge discussions with B.-Y. Ha,
Y. Levin, R. van Roij and P. B. Warren.

\section*{Appendix}

Rather generally, when we are evaluating a configurational integral,
with the aim of calculating the free energy,
we can split the variables over which we are integrating into two sets.
Then integrating over one set is always possible in principle,
and it results in an effective Hamiltonian
which is a function only of the remaining set of variables.
However, this effective Hamiltonian is not in general
pairwise additive, simply because there is no reason for it to be.
If we denote the set of all variables by $v^{N+M}$, and the two subsets
by $v^N$ and $v^M$ then the configurational integral is
\begin{equation}
Z=\int {\rm d}v^{N+M}\exp\left(-H(v^{N+M})/kT\right)=
\int {\rm d}v^N\exp\left(-H'(v^N)/kT\right),
\end{equation}
where
\begin{equation}
\exp\left(-H'(v^N)/kT\right)=\int {\rm d}v^M\exp\left(-H(v^{N+M})/kT\right).
\label{hdash}
\end{equation}
$H'$ will not in general be expressible as a sum over a pair potential
even if $H$ can be.
Physically, this is most often useful when the
two sets of coordinates, $v^N$ and $v^M$, are very different.
We consider a model in which the set of coordinates $v^M$
is coarse grained \cite{chaikin},
i.e., the integration over $v^M$ is left as a function of a
set of coarse-grained variables,
so instead of Eq. (\ref{hdash}) we have,
\begin{equation}
\exp\left(-H'(v^N,c^L))/kT\right)=
\int {\rm d}v^M
\exp\left(-H(v^{N+M})/kT\right)
\prod_{a=1}^L\delta\left(c_a(v^M)-c_a\right),
\end{equation}
where $c^L$ is a set of $L\ll M$ coarse-grained variables.
$c_a$ is the $a$th coarse-grained variable and
$c_a(v^M)$ is the definition of the $a$th coarse-grained
variable in terms of the microscopic coordinates, $v^M$.
The configurational integral is then
\begin{equation}
Z=\int {\rm d}v^N{\rm d}c^L
\exp\left(-H'(v^N,c^L)/kT\right).
\end{equation}
Our model has a configurational integral, Eq. (\ref{zn}), of this form,
with $L=N$ as there is one coarse-grained variable per particle.
The $v^N$ and $c^N$ coordinates are the positions
and $s$ variables of the particles.

%\newpage

%\newpage

%\begin{figure}
%\caption{
%The phase diagram of hard spheres plus a long-range
%long-range attraction which has two parts, one independent of the
%fluctuating internal variable $s$ and the other
%proportional to $s$.
%$\nu_0<0$, $\nu_1=\nu_0/2$ and the reduced temperature
%$T^*=kT\alpha\sigma^6/\nu_1^2$.
%See caption to Fig. \ref{fig1} for the meaning of the letters, curves and
%horizontal lines.
%}
%\label{fig2}
%\begin{center}
%\epsfig{file=fig2.eps,width=4.0in}
%\end{center}
%\end{figure}

%\newpage

%\newpage

%\newpage

\end{document}